\documentclass[preprint2,longabstract]{aastex}

\usepackage{latexsym,wasysym}

\def \al26{\mbox{$^{26}$Al}}
\def \mg26{\mbox{$^{26}$Mg}}
\def \a27{\mbox{$^{27}$Al}}
\def \fe60{\mbox{$^{60}$Fe}}
\def \ni60{\mbox{$^{60}$Ni}}
\def \n62{\mbox{$^{62}$Ni}}
\def \f56{\mbox{$^{56}$Fe}}
\def \cl36{\mbox{$^{36}$Cl}}
\def \calcium41{\mbox{$^{41}$Ca}}
\def \palladium107{\mbox{$^{107}$Pd}}
\def \chl35{\mbox{$^{37}$Cl}}
\def \ca40{\mbox{$^{40}$Ca}}
\def \pd107{\mbox{$^{107}$Pd}}
\def \be10{\mbox{$^{10}$Be}}
\def \beryl9{\mbox{$^{9}$Be}}
\def \msun{\mbox{M$_{\odot}$}}
\def \o16{\mbox{$^{16}$O}}
\def \ox17{\mbox{$^{17}$O}}
\def \oxy18{\mbox{$^{18}$O}}
\def \s36{\mbox{$^{36}$S}}
\def \mn53{\mbox{$^{53}$Mn}}
\def \ch53{\mbox{$^{53}$Cr}}
\def \hf182{\mbox{$^{182}$Hf}}
\def \zr96{\mbox{$^{96}$Zr}}

\slugcomment{Submitted to The Astrophysical Journal}

\shorttitle{\al26 in planetary systems} 
\shortauthors{Gaidos et al.}

\begin{document}

\title{\al26 and the Formation of the Solar System from a Molecular Cloud Contaminated by Wolf-Rayet Winds}

\author{Eric Gaidos\altaffilmark{1}} \affil{Department of Geology and
Geophysics, University of Hawaii, Honolulu, HI 96822}
\email{gaidos@hawaii.edu}

\author{Alexander N. Krot\altaffilmark{1}} \affil{Hawaii Institute of
Geophysics and Planetology, University of Hawaii, Honolulu, HI 96822}
\email{sasha@higp.hawaii.edu}

\author{Jonathan P. Williams\altaffilmark{1}} \affil{Institute for
Astronomy, University of Hawaii, Honolulu, HI 96822}
\email{jpw@ifa.hawaii.edu}

\and

\author{Sean N. Raymond\altaffilmark{1}}
\affil{NASA Postdoctoral Program Fellow, Center for Astrophysics and
Space Astronomy, University of Colorado, Boulder, CO 80309} 
\email{raymond@lasp.colorado.edu}

\altaffiltext{1}{NASA Astrobiology Institute}

\begin{abstract}

In agreement with previous work, we show that the presence of the
short-lived radionuclide \al26 in the early Solar System was unlikely
($<2$\% {\it a priori} probability) to be the result of direct
introduction of supernova ejecta into the gaseous disk during the
Class II stage of protosolar evolution.  We also show that Bondi-Hoyle
accretion of any contaminated residual gas from the Sun's natal star
cluster contributed negligible \al26 to the primordial Solar System.
Our calculations are consistent with the absence of the oxygen
isotopic signature expected with any late introduction of supernova
ejecta into the protoplanetary disk.  Instead, the presence of \al26
in the oldest Solar System solids (calcium-aluminum-rich inclusions or
CAIs) and its apparent uniform distribution with the inferred
canonical \al26/\a27 ratio of $(4.5-5) \times 10^{-5}$ support the
inheritance of \al26 from the Sun's parent giant molecular cloud.  We
propose that this radionuclide originated in a prior generation of
massive stars that formed in the same molecular cloud and contaminated
that cloud by Wolf-Rayet winds.  We calculated the Galactic
distribution of \al26/\a27 ratios that arise from such contamination
using the established embedded cluster mass and stellar initial mass
functions, published nucleosynthetic yields from the winds of massive
stars, and by assuming rapid and uniform mixing into the cloud.
Although our model predicts that the majority of stellar systems
contain no \al26 from massive stars, and that the {\it a priori}
probability that the \al26/\a27 ratio will reach or exceed the
canonical Solar System value is only $\sim$6\%, the maximum in the
distribution of {\it non-zero} values is close to the canonical
\al26/\a27 ratio.  We find that the Sun most likely formed 4-5 million
years (Myr) after the massive stars that were the source of \al26.
Furthermore, our model can explain the initial Solar System abundance
of a second, co-occuring short-lived radionuclide, \calcium41, if
$\sim 5 \times 10^5$~yr elapsed between ejection of the radionuclides
and the formation of CAIs.  The presence of a third radionuclide,
\fe60, can be quantitatively explained if (a) the Sun formed
immediately after the first supernovae from the earlier generation of
stars; (b) only 5\% of supernova ejecta was incorporated into the
molecular cloud , or (c) the radionuclide originated in an even
earlier generation of stars whose contributions to other radionuclides
with a shorter half-life had completely decayed.
\end{abstract}

\keywords{planetary   systems:  protoplanetary  disks   ---  planetary
systems: formation}

\section{Introduction}

\subsection{Short-lived radionuclides in the early Solar System}

Primitive meteoritic materials contain compelling evidence for the
short-lived radionucides (SLRs) \be10, \al26, \cl36, \calcium41,
\mn53, \fe60, \pd107, and \hf182 in the early Solar System
\citep{Goswami05}.  These radionuclides have a half-life $\tau_{1/2} <
10$~Myr and are potential high-resolution chronometers of events
during the epoch of planet formation \citep{Kita05,Krot08a}, but only
if they were introduced at discrete times and were uniformaly
distributed in the Solar System.  The potential sources of these
radionuclides inform us about the young Sun's stellar neighborhood
\citep{Hester04}, or its magnetic interaction with an accretion disk
\citep{Shu97}.  The decay of one short-lived radionuclide, \al26,
might have been the principle heat source in planetesimals and
responsible for the differentiation of the parent bodies of magmatic
iron meteorites in the first 1-2 Myr of the Solar System
\citep{Greenwood05,Schersten06,Markowski06}.

The origin of SLRs is controversial.  The half-life of each is much
shorter than the $\sim$100~Myr mixing time of the interstellar medium
\citep{deAvillez02} and the excess abundances of at least five
radionuclides (\be10, \cl36, \al26, \calcium41, and \fe60) require one
or more ``local'' sources in addition to the average Galactic
background \citep{Jacobsen05}.  Two principle scenarios emerged soon
after the first reports of fossil SLRs in meteorites: ({\it i}) an
origin in one or more neighboring massive stars, either Type II
supernova (SN) progenitors \citep{Cameron77} or Wolf-Rayet (WR) stars
\citep{Arnould86}; and ({\it ii}) production by irradiation of gas or
dust with energetic particles from the active young Sun
\citep{Heymann76}.  An alternative scenario invoking an origin in a
nearby intermediate-mass asymptotic giant branch star
\citep{Wasserburg94} is generally discounted because of the very low
probability of such an encounter \citep{Kastner94}.  Each of the two
schools of thought has developed elaborate models
\citep{Lee98,Gounelle06,Ouellette07} but neither has produced a
comprehensive explanation of the origin, abundance, and distribution
of all the SLRs \citep{Goswami05,Gounelle06,Duprat07}.

The origins of three SLRs seem unambiguous: ({\it i}) $^{10}$Be
($\tau_{1/2} \sim 1.5$~Myr) inferred from excess $^{10}$B
\cite{McKeegan00} could have been produced by energetic particle
irradiation but not by stellar nucleosynthesis.  Although magnetic
trapping of Galactic cosmic rays in the protosolar molecular cloud has
been advanced as an alternative explanation \citep{Deschetal04}, it is
inconsistent with variations in the inferred initial \be10/\beryl9
ratio \citep{McKeegan00,Sugiura01,Marhas02,MacPherson03}.  ({\it ii})
Excess \s36 correlated with the ratio $^{35}$Cl/$^{34}$S and
attributable to the decay of \cl36 ($\tau_{1/2} \sim 0.3$~Myr) was
reported in sodalite, an alteration phase that replaced anorthite in
CAIs and chondrules from CV chondrites \citep{Lin05,Hsu06}.  The
inferred \cl36/$^{35}$Cl ratio at the time of sodalite formation is
$5\times10^{-6}$.  If the sodalite formed late ($>$1.5~Myr after CAI
crystallization, based on absence of \al26), the initial
\cl36/$^{35}$Cl ratio was $>1.6\times 10^{-4}$.  This is inconsistent
with a massive stellar source and requires a late episode of
irradiation.  ({\it iii}) $^{60}$Ni excess correlated with the
$^{56}$Fe/$^{58}$Ni ratio is evidence for live \fe60, a radionuclide
that cannot be produced by irradiation and must have originated in one
or more massive stars \citep{Tachibana03}.  High-precision nickel
isotope measurements in several groups of magmatic iron meteorites
indicate that \fe60 was uniformly distributed in the solar nebula
\citep{Dauphas08} but its initial abundance, (\fe60/\f56)$_0 \sim
(0.5-1)\times 10^{-6}$, is uncertain
\citep{Tachibana06,Bizzarro07,Quitte07,Guan07}.  The lower end of the
range of estimates is consistent with the expected abundance in
star-forming regions if star formation rates were approximately twice
as high at the epoch of Solar System formation
\citep{Williams08,Gounelle08a} and/or the half-life of \fe60 is
actually longer than the published value ($\tau_{1/2} \approx
1.5$~Myr).

\subsection{The origin of \al26}

Ironically, the origin of \al26 ($\tau_{1/2} \sim 0.73$~Myr), the
first SLR to be discovered \citep{Lee76} and the one best studied,
remains an enigma.  A significant contribution by irradiation is
disputed on the grounds that ({\it i}) production models adjusted to
achieve the required \al26/\a27 ratio over-predict the observed
initial abundance of $^{41}$Ca ($\tau_{1/2} \sim 0.1$~Myr) whose
co-occurrence indicates a common origin \cite{Sahijpal98,Goswami05},
({\it ii}) the flux of energetic particle inferred from X-ray
observations of young solar-type stars would have been insufficient to
produce the \al26 \cite{Duprat07}, and ({\it iii}) a lack of
correlation between the presence of \al26 and \be10, which {\it was}
formed by irradiation \citep{MacPherson03,Marhas04}.  Finally, the
canonical value [$(4.5-5)\times10^{-5}$] of the initial \al26/\a27 ratio
in the majority of CAIs from primitive chondrites
\citep{MacPherson95,Jacobsen08,Makide08}, the consistent chronology of
CAI and chondrule formation between $^{207}$Pb-$^{206}$Pb and
\al26-$^{26}$Mg systematics \citep{Amelin02,Halliday06,Connelly08},
and the apparently uniform Mg-isotope compositions of bulk chondrites,
Mars, Moon and the Earth \citep{Thrane06} are evidence for a uniform
distribution thought inconsistent with a central (solar) irradiation
source.

Although these observations favor an origin of \al26 in massive stars,
the mechanism and timing of its delivery to the early Solar System
remains unclear.  Two models have been proposed: instability-induced
mixing of gaseous SN ejecta during the molecular cloud core phase
\citep{Cameron77}, and injection of SLR-bearing dust grains into the
later protoplanetary disk \citep{Ouellette05}.  The relatively small
cross-section of the solar nebula dictated that the progenitor was
within $\sim$1~pc of the Solar System \citep{Looney06}.  Such a
circumstance is possible only in the dense environment of a large
stellar cluster \citep{Hester04}.  Incorporation of hot, low-density
gas from SN ejecta into the denser, cooler disk gas is inefficient
\cite{Vanhala02,Ouellette07,Boss08}.  Instead, Ouellette et al. (2005)
proposed that several SLRs, including \al26, were delivered to the
early Solar System as grains that condensed from SN ejecta and
vaporized upon entering the relatively high-density gas in the disk
\citep{Ouellette07}.  Disks around low-mass cluster stars persist for
up to 6~Myr \citep{Haisch05,Jayawardhana06}, longer than the main
sequence lifetime of the most massive SN progenitors
\citep{Schaller92}.

However, any explanation for Solar System \al26 invoking a late
introduction of SN ejecta has four significant shortcomings : ({\it i})
Not all stars form in large clusters and most clusters are dynamically
unbound and disperse in $\sim$10~Myr \citep{Lada03}.  It is
statistically unlikely that the Sun would have been suffiently close
to a massive star at the end of the latter's main sequence life
\cite{Williams07,Gounelle08b}.  ({\it ii}) Even the most massive stars
have a main sequence lifetime of at least 3~Myr and, if they and the
Sun formed simultaneously, the former would have ended in supernovae
late in the evolution of the solar nebula and probably long after CAIs
containing the canonical \al26/\a27 ratio had formed.  These CAIs, the
oldest dated solids from the Solar System \cite{Amelin02}, have a
narrow range of inferred initial \al26/\a27 ratios suggesting that
they formed in $\ll$1~Myr \citep{Thrane06,Jacobsen08}, consistent with
the duration of Class 0-I stages of protostars
\citep{Smith04,Ward-Thompson07}.  A short interval of CAI formation is
also consistent with the narrow range of their oxygen isotope
compositions ($\Delta^{17}{\rm O} = -24\pm 2 \permil$)
\citep{MacPherson08,Makide08}, which are similar to the inferred
oxygen isotopic composition of the Sun \citep{Hashizume05}.  [Later
  CAI formation would presumably have reflected the rapid oxygen
  isotopic evolution of the solar nebula along the slope-one
  carbonaceous chondrite anhydrous minerals (CCAM) line towards the
  terrestrial value ($\Delta^{17}{\rm O} = 0\permil$), a trend
  attributed to CO photochemical self-shielding and radial mixing in
  the disk \citep{Yurimoto07,Aleon07}.]  ({\it iii}) Late injection of
  \al26 into the protoplanetary disk would likely have disturbed the
  oxygen isotopic composition of the disk from the CCAM line, and this
  is not observed \citep{Gounelle07}.  ({\it iv}) SN models that
  produce the canonical \al26/\a27 ratio invariably over-predict the
  abundance of \mn53 and must impose fallback of the innermost layers,
  e.g., Meyer (2005).  (These models also over-predict the abundance
  of \fe60, see \S \ref{sec.others}).

A scenario in which SLRs from coeval SN progenitors were injected into
the protoplanetary disk can be evaluated for its statistical
plausibility.  Gounelle \& Meibom (2007) estimated that the
probability that any given disk is contaminated by a SN with enough
\al26 to reach the canonical \al26/\a27 ratio is $\le 0.3$\%.
However, there are three limitations to their model: ({\it i}) It
underestimated the probability by assuming a maximum disk radius of
50~AU based on observations and theoretical arguments that disks
within 0.1~pc of massive (O) stars suffer from photoevaporation
\cite{Johnstone98,Chevalier00}.  But disks further from O stars are
invariably larger \citep{Vicente05,Andrews07,Balog07} and this effect
compensates the greater distance from the source of radionuclides.
({\it ii}) Their model overestimated the probability by neglecting the
expansion of the host star cluster and using the $\le$1~Myr-old Orion
nebular cluster as a template, rather than a cluster at the minimum
age (3~Myr) when supernovae occur.  In older clusters, stars are more
widely separated.  ({\it iii}) It only included the contribution of SN
to \al26.  \al26 is also produced and ejected from stars with initial
masses $\ge 40$~\msun~in their Luminous Blue Variable (LBV) and
Wolf-Rayet (WR) phases \cite{Arnould86,Palacios05,Sahijpal06}.

\subsection{Was Solar System  \al26 inherited?}

The uniform distribution of \al26, the existence of the canonical
\al26/\a27 ratio in CAIs, the adherence of primitive oxygen isotope
compositions to the CCAM line, and the low likelihood of a SN
injection event all point to the introduction of \al26 before the
collapse of the protosolar cloud.  Indeed, the oxygen isotopic
composition of the {\it entire} Solar System appears to be displaced
from the locus of mean Galactic evolution \citep{Young08}, suggesting
primordial contamination.  The source of \al26 must also have
introduced \calcium41, which co-occurs with an inferred initial
abundance of \calcium41/\ca40 $\sim1.5 \times 10^{-8}$
\citep{Sahijpal98}.  The half-life of this radionuclide is only
0.1~Myr, limiting any time delay between production at the source and
its incorporation into CAIs.  The source of \al26 cannot have been
accompanied by substantial \mn53, as is the case of SN ejecta without
fallback onto the remnant \citep{Meyer05}.  (The predicted \mn53
abundance in the interstellar medium is consistent with its inferred
initial abundance in the Solar System and a ``local'' source is not
required.)  Any relationship between the source of \al26 and that of
\fe60 remains to be determined.  There is as yet no evidence that
\fe60 and \al26 are correlated and had the same origin.  Indeed, there
may be a {\it deficit} of \fe60 relative to \al26 compared to SN
ejecta that cannot be explained by free decay of the two
radionuclides: The initial ratio of \fe60 to \al26 in the Solar System
was 0.1-0.2 \citep{Tachibana06}, lower than the 0.3 deduced from the
Galactic average $\gamma$-ray emission \citep{Wang07} and theoretical
predictions \citep{Limongi06,Woosley07}.

\begin{figure}
\epsscale{1.0}
\plotone{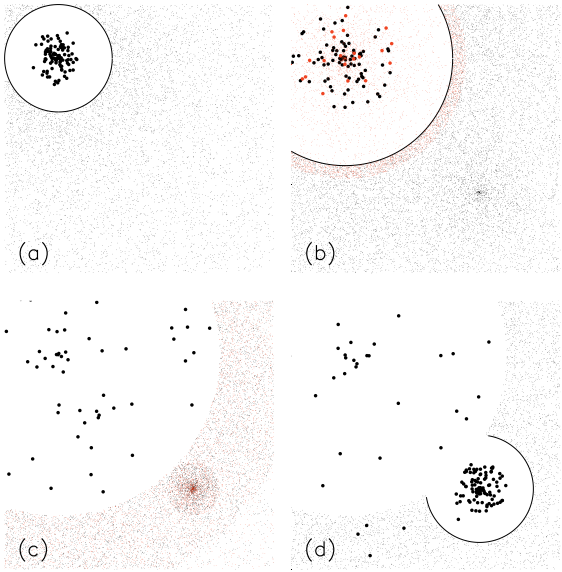}
\caption{Cartoon of the molecular cloud contamination scenario to
explain the presence of \al26 in the early Solar System: (a) A young
cluster of stars forms an HII region in a giant molecular cloud; (b)
massive stars eject \al26 into the cloud; (c) a clump collapses from
the contaminated cloud; (d) a second generation of stars (including
the Sun) forms from the clump.}
\label{fig.cartoon}
\end{figure}

Very massive ($\ge 40$~\msun) stars eject \al26 (and other SLRs)
during the Wolf-Rayet phase of mass loss near the end of hydrogen core
burning, as well as in SN \citep{Arnould86}.  WR winds might account
for a large fraction of the total fluence and Galactic distribution of
$\gamma$-rays from the decay of \al26
\cite{Palacios05,Diehl06,Voss08,Martin08}.  [The non-detection of
  $\gamma$-ray emission in the decay line of \al26 from the nearest WR
  star $\gamma^2$-Velorum can be explained by the dispersal of most of
  the radionuclide to large angular separation \citep{Mowlavi06}].  We
propose that most or all of the \al26 in the early Solar System
originated in WR winds from one or more massive stars that
contaminated the molecular cloud from which the Sun formed.  (Sahijpal
\& Soni (2006) also considered the contribution of WR winds to Solar
System inventories of SLRs.)  These stars could have been members of
the same embedded cluster as the Sun, or, more likely, members of
another cluster that formed in the same giant molecular cloud (GMC)
(Figure \ref{fig.cartoon}).  An analogous ``self-contamination''
scenario has been invoked to explain anomalous abundance patterns in
some globular clusters \citep{Smith06}.

Our proposal is based upon the following: ({\it i}) The amount of
\al26 ejected in the winds of a single 60 \msun~star \cite{Limongi06}
is sufficient to contaminate $2 \times 10^4$ \msun~of
solar-metallicity gas to the canonical \al26/\a27 ratio of the Solar
System. ({\it ii}) The WR phase can occur as soon as 1-2~Myr after
star formation and more than 1~Myr before the SN \citep{Palacios05},
making it more likely to contaminate residual star-forming molecular
gas.  ({\it iii}) WR winds have speeds of up to 2000~km~sec$^{-1}$
\citep{Niedzielski02} and can traverse star-forming regions
(10-100~pc) in $10^4-10^5$~yr.  ({\it iv}) In contrast to single
clusters where star-formation may be co-eval, star formation in a
molecular cloud can occur over an interval of at least a few Myr
\citep{Hartmannetal01}, and possibly longer \citep{Williams00} as
clumps of gas with a mass spectrum $M^{-1.7\pm0.1}$ \citep{Pudritz02}
form multiple embedded stellar clusters \citep{Williams00}.  For
example, the Orion star-forming complex contains several subgroups
that are several Myr older than the Orion nebula cluster
\citep{Bally08}({\it v}) Wolf-Rayet winds contain multiple SLRs,
including \al26, \calcium41, and \cl36, but little or no \mn53 or
\fe60 \citep{Arnould06}.  We propose that the collapse of the
protosolar cloud homogenized the distribution of these isotopes (but
see \S \ref{sec.funcais}).  Our scenario does not preclude a
SN-triggered collapse \citep{Cameron77,Boss08}, which would have
occured {\it after} the WR phase.

In \S \ref{sec.disk} we revisit the scenario of \al26-introduction by
SN into the protoplanetary disk with a Monte Carlo approach that used
more realistic disk sizes, accounted for the expansion of clusters,
and included the contribution from both WR winds and SN.  While our
results differ quantitatively from those of Gounelle \& Meibom (2007),
the calculated probability of a disk having the canonical Solar System
\al26/\a27 ratio is nevertheless small ($<$2\%).  We also consider
Bondi-Hoyle accretion of contaminated intracluster gas, as proposed by
Throop \& Bally (2008) and find its inclusion does not significantly
alter this result (\S \ref{sec.bondihoyle}).  We then use a similar
Monte Carlo model to investigate a scenario where \al26 is
introduced by WR winds into the parent molecular cloud of the Sun (\S
\ref{sec.cloud}).  This model readily reproduces the canonical
\al26/\a27 ratio of the Solar System.  We discuss the delivery of
\al26 from WR winds into the molecular cloud, interpret CAIs that lack
the canonical \al26/\a27 ratio, and, in the context of the cloud
contamination scenario, present additional calculations for three
other SLRs (\calcium41, \fe60, and \cl36) in \S \ref{sec.discussion}.

\section{Disk contamination scenario\label{sec.disk}}

We calculated the Galactic distribution of \al26/\a27 ratio in the
disk injection scenario \citep{Ouellette05} by extending the model of
Williams \& Gaidos (2007) to include \al26 from both winds and SN, a
dynamically realistic description of cluster expansion, and the effect
of the UV field of massive stars on disk size.  Unlike Williams \&
Gaidos (2007) we do not explicitly consider disk evolution and
disappearance in our model because this occurs on a timescale of
$\sim$6~Myr \citep{Haisch05}, much longer than the likely epoch of CAI
formation.

For simplicity, we assume that every young low-mass star has a disk
and that the properties of the disk are independent of stellar mass.
According to Lada \& Lada (2003), 72\% of stars form in clusters with
$N_* > 100$ members distributed in size according to $dN_c/dN_*\propto
N_*^{-2}$.  The remaining 28\% form in isolation or in clusters with
fewer than $N_* = 100$ stars.  The most massive members of such small
clusters will be $\sim$5~\msun~B stars that do not produce winds or
Type II core-collapse SN.  Disks around members of such clusters will
not receive any exogenous \al26.  The other 72\% of disks were
represented by $10^5$ Monte Carlo calculations.  The size of the host
cluster of each disk was drawn from a $N_*^{-1}$ distribution ($N_*
\ge 100$).  The number of massive stars (SN progenitors with $M_* >
8$\msun) in the host cluster was selected from a Poisson distribution
with an average of $3 \times10^{-3}N_*$ \citep{Williams07}.  The
masses of these stars were drawn from a power-law initial mass
function $dN_*/dM_*\propto M_*^{-2.5}$ \cite{Kroupa02}.  Clusters form
over $\sim$1~Myr \citep{Hillenbrand07} and most star formation in a
single cluster occurs within $\sim$3~Myr
\citep{Hartmann01,Hartmann03,Huff06,Jeffries07,Hillenbrand07}.  We
assumed the instantaneous formation of all massive stars and an
exponentially decaying rate of low-mass star formation after massive
star formation.  We used age statistics for members of the Orion
Nebula Cluster \citep{Palla99} to infer a decay time of 2.7~Myr, which
we adopted for clusters of all sizes.  We assumed that the rate of
star formation does not depend on stellar mass.

The \al26/\a27 ratio in the disk around the $i$-th low-mass star at an
interval $T$ after the star's formation at time $t^*_i$ was calculated
by summing over the product of the the yield $m_j$ of the radionuclide
from the $j$-th wind or SN, the solid angle subtended by the disk at
the time of ejection $t_j$, and the factor of free decay between that
time and $t^*_i + T$;
\begin{equation}
\label{eqn.al26}
(\al26/\a27)_i = \frac{r_d^2}{8m_df_{Al}}\Sigma_{j}
\frac{m_{j}}{d_{ij}^2}exp \left[\log 2 (t_{j} - t^*_i -
T)/\tau_{1/2}\right],
\end{equation}
where $r_d$ and $m_d$ are the (constant) radius and mass of the disk,
$f_{Al}$ is the mass fraction of \a27, and $d_{ij}$ is the distance of
the disk from the source star during the at the time of ejection.  We
considered injection as instantaneous because the speeds of SN ejecta
and winds are $>1000$~km~s$^{-1}$, and the \al26-producing WR phase of
an individual massive star is brief ($\sim$1~Myr) compared to the
dispersal timescale of the cluster ($\sim$10~Myr).  To account for the
changing perspective of each star as it orbits inside the cluster, we
used the isotropic average of the projected cross-section of each
disk, $\pi r_d^2/2$.  $m_j$ and $t_j$ were estimated by spline
interpolation in a grid of yield calculations by Limongi \& Chieffi
(2006).  We used a default disk radius of 200~AU
\citep{Vicente05,Andrews07} but to account for photoevaporation by the
UV radiation from massive stars \citep{Johnstone98} we reduced this to
30~AU for disks within 0.2~pc of the cluster center at 3~Myr, and to
zero for disks within 0.1~pc.  This was a conservative assumption,
since the evidence for significant disk truncation is weak
\citep{Balog07}, but it only has a minor effect on our results.  Like
Ouellette et al. (2005), we adopted a Minimum Mass Solar Nebula disk
mass of 0.013~\msun~and the fractional abundance of \a27 given
by Lodders (2003).

As in Williams \& Gaidos (2007), we placed the massive stars at the
cluster center \cite{Grebel07} and assumed that, at any time, low-mass
stars were distributed with an inverse-square density profile, such
that an equal number of stars reside in shells of constant thickness
out to the edge of the cluster at $r_c$.  To model cluster expansion,
and thereby determine the distance of a disk to each source of \al26,
we developed an empirical relationship for the time dependence of
$r_c$ based on a series of numerical simulations of clusters
containing between 100 and 15,000 stars.  The dynamical simulations
were preformed using the NBODY4 code running on the Cambridge
University GRAPE-6a card \cite{Aarseth03}.  In each case the stars
were initially (3~Myr) distributed in a Plummer sphere
\citep{Binney88} with a radius set by the requirement that the initial
surface density $\Sigma_3 = 100$~pc$^{-2}$ \citep{Williams07}.  The
virial parameter $\Omega$ (ratio of kinetic to gravitational potential
energy) was set to 1.5.  This condition is brought about by a cluster
local star formation efficiency of 33\% in the embedded cluster and
the instantaneous removal of the remaining gas at 3~Myr
\citep{Bastian06}.  The size of the cluster was explicitly determined
at regular intervals until an age of 10~Myr.  We found that the
expansion of the cluster from its size at 3~Myr was closely
approximated by
\begin{equation}
\label{eqn.exp}
r_c(t) \approx r_c(3~{\rm Myr}) + 0.45\left[2G\bar{M_*}\Omega\sqrt{\pi \Sigma_3
N_*}\right]^{1/2}(t-3~{\rm Myr}),
\end{equation}
where $G$ is the gravitational constant and $\bar{M_*}$ is the average
stellar mass.  The expression multiplying 0.45 is the cluster's virial
speed \citep{Binney88}.  In each of our simulations the surface
density of the cluster fell to the background level of field stars
(2-3~pc$^{-2}$) by 10~Myr, in agreement with observations
\citep{Lada03}.

Eqn. \ref{eqn.exp} specifies the radius of the cluster at the epoch
$t_j$ of a massive stellar wind or SN.  The uniform distribution of
low-mass stars with distance from the cluster center $0 < d_{ij} <
r_c(t_j)$ was then used in Eqn. \ref{eqn.al26} to produce a
distribution of \al26/\a27.  These distributions were summed over all
events in a cluster, corrected for free decay, averaged over $10^5$
realizations, and multipled by 0.78 to produce a Galactic
distribution.

\begin{figure}
\epsscale{1}
\plotone{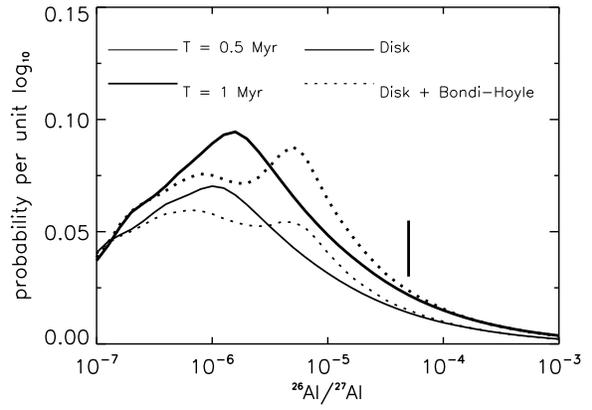}
\caption{The distribution of non-zero values of the \al26/\a27 ratio
predicted by a Monte Carlo model of its formation in massive stars and
incorporation injection into the protoplanetary disk (see text for
details).  The units of the ordinate are fractional number of Monte
Carlo systems per unit common logarithm.  The vertical bar demarks the
canonical Solar System value of $5 \times 10^{-5}$.  The curves do not
integrate to unity because a large majority of systems are not
contaminated with \al26 (Table \ref{tab.stats}).  The solid curves are
for the disk injection scenario (see text), and the dashed curves
include Bondi-Hoyle accretion of nebular gas.  Two values for the
elapsed time between the formation of the Sun and \al26-containing
CAIs are considered.}
\label{fig.disk}
\end{figure}

The calculated distributions of \al26/\a27 are plotted in Figure
\ref{fig.disk} and the fractions of systems that have any initial
\al26 or \al26 abundances exceeding the canonical value are given in
Table \ref{tab.stats}.  We also report the 95-percentile values of
\al26/\a27.  We considered two values for $T$, which in the Solar
System represents the epoch of CAI formation.  The probability is
1.1\% for $T=0.5$~Myr, rising to 1.9\% by 1~Myr.  The probability is
higher at still later, but unlikely CAI formation times (not shown).
Our probablities are several times higher than that reported by
Gounelle \& Meibom (2008), mostly due to the larger disk size we used,
but are still small.  Our estimates are nonetheless optimistic because
we assume all high-mass stars form prior to low-mass stars.  If all
stars form instantaneously, then nearly 3~Myr must elapse before \al26
is produced, and {\it no} disks contain \al26 by the time of CAI
formation.

\section{Bondi-Hoyle accretion scenario \label{sec.bondihoyle}}

Disks might also accrete gas from their natal cluster before it is
removed by winds and SN explosions.  Throop \& Bally (2008) proposed
that Bondi-Hoyle accretion of residual, SN-contaminated cluster gas
onto the protosolar nebula produced isotopic anomalies in the Solar
System, including the presence of SLRs.  They estimated that disks
around solar-mass stars in a 3000-star cluster could accrete an
additional $\sim$0.01~\msun~of gas, an amount comparable to the
original mass of a disk, over the 2-4~Myr that gas remained in the
cluster.  We estimated the amount of \al26 that could be introduced by
this process by specifying the fraction of disk mass $f$ acquired by
Bondi-Hoyle accretion by the time of CAI formation, and assuming that
accretion is constant as long as intracluster gas is present.  We
assumed that the maximum amount of mass that can be accreted by a disk
is equal to the initial disk mass \citep{Throop08}, i.e. $f \leq 0.5$.
To account for the absence of cluster gas during the history of
later-forming stars we adjusted the accreted mass by the ratio of the
time interval between low-mass star formation and the disappearance of
cluster gas, and the lifetime of the cluster gas.  Like Throop \&
Bally (2008), we assumed a gas lifetime of 3~Myr.  We calculated the
average \al26/\a27 ratio of cluster gas during the period of
Bondi-Hoyle accretion (see \S \ref{sec.cloud} for details) and then
determined the final \al26/\a27 of the disk as \al26/\a27 =
(1-f)(\al26/\a27)$_0$ + f(\al26/\a27)$_{acc}$ where the subscripts
refer to the initial value and the average value during accretion.

We calculated distributions of the \al26/\a27 ratio for combined
Bondi-Hoyle accretion and disk injection and found that the former has
a negligible effect (Figure \ref{fig.disk} and Table \ref{tab.stats}).
This is because low-mass stars that form late enough ($\sim$3~Myr) to
acquire significant \al26 will accrete little gas because the
intracluster gas disappears soon thereafter.  This is of course
entirely a result of our (reasonable) assumption that intracluster gas
is evacuated by the time the massive stars leave the main sequence, if
not earlier.  Nevertheless the same cosmochemical timing arguments that
apply to the disk injection scenario also apply to the Bondi-Hoyle
scenario; CAIs probably formed by the time a disk had formed, or very
soon thereafter, and thus later accretion of contaminated gas cannot
be responsible for the presence of \al26 in them.

\section{Molecular cloud contamination scenario \label{sec.cloud}}

In this scenario the Sun's natal giant molecular cloud spawned an
earlier generation of massive stars [Figure \ref{fig.cartoon}(a)]
whose WR winds contaminated the rest of the cloud (b), from which the
Sun subsequently formed, perhaps in a second cluster (c and d).  We
calculated the distribution of \al26/\a27 ratios in this scenario by
$10^5$ Monte Carlo simulations, each corresponding to a disk formed
from a GMC contaminated by an immediately previous generation of
massive cluster stars.  The number and masses of those stars were
drawn from the distributions described in \S \ref{sec.disk}.  The mass
of gas in the GMC was calculated using an average stellar mass derived
from the initial mass function of Kroupa (2002), and a total
star-formation efficiency of 10\% \cite{Williams00}.  The amount of
\al26 added to the molecular cloud by massive stellar winds was
calculated as a function of time using the yields and times of Limongi
\& Chieffi (2006) and by assuming 100\% delivery efficiency and
instantaneous mixing into the cloud (we discuss this assumption
further in \S \ref{sec.efficiency}).  We did not include SN ejecta in
these calculations, but include it when we consider \fe60 in \S
\ref{sec.others} (but see the footnote on SN ejecta delivery in \S
\ref{sec.efficiency}). We assumed that the Sun and CAIs formed
simultaneously.

\begin{figure}
\epsscale{1}
\plotone{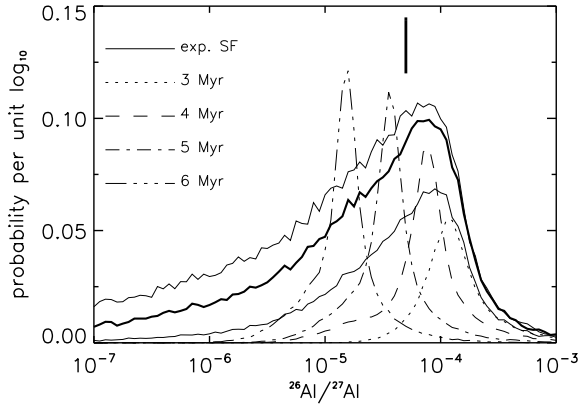}
\caption{The distribution of non-zero values of the \al26/\a27 ratio
predicted by an alternative model in which \al26 was produced in an
earlier generation of massive stars and introduced by Wolf-Rayet winds
into the molecular cloud that formed the Sun (see text for details).
The units of the ordinate are fractional number of Monte Carlo systems
per unit common logarithm.  The vertical bar is the canonical Solar
System value of $5 \times 10^{-5}$.  The integral of the curves are
not unity because the majority of systems are not contaminated with
\al26 (Table \ref{tab.stats}).  The broken curves are for a monotonic
elapsed time between the formation of the earlier generation of
massive, \al26-producing stars and the Sun.  The solid curves (shown
one-tenth scale) are for an exponentially-decaying rate of star
formation with a decay time of 2.7~Myr (heavy curve) or 1.7 and
3.7~Myr (light curves).  The area under the solid curves changes with
assume star formation history, but the location of the peak does not.}

\label{fig.cloud}
\end{figure}

We first carried out a series of calculations for different intervals
of elapsed time (3-6~Myr) between the formation of the earlier
generation of massive stars and the Sun (dashed lines in Figure
\ref{fig.cloud}).  If the interval is less than 3~Myr contamination
has yet to take place in our model and newly-formed stellar systems
lack \al26.  An interval of 4-5~Myr is most likely to produce the CAI
value.  The history of low-mass star formation in molecular cloud
complexes is poorly known but clearly non-monotonic
\citep{Hartmannetal01,Hartmann03}.  We calculated more realistic
distributions by again adopting the exponential rate with a 2.7~Myr
decay time based on the data of Palla \& Stahler (1999).  This is
plotted as the heavy solid line in Figure \ref{fig.cloud}.  That curve
can be understood as a convolution of the star formation history with
the \al26/\a27 distributions for ``starburst'' scenarios.  The
probability of \al26/\a27 exceeding the canonical value is 6.2\%.
This figure depends on the assumed star formation history: For
example, varying the decay time constant by $\pm$1~Myr changes the
fraction between 4.3 and 6.5\%.  However, the peaks in all three
distributions are near the CAI value (solid lines in Figure
\ref{fig.cloud}).  This robustness is a result of negligible \al26
production at times earlier than 3~Myr and negligible low-mass star
formation at times much later than 5~Myr.  We found that simulations
which produce a \al26/\a27 ratio within 2$\sigma$ [$\sigma = 0.1
  \times 10^{-5}$ \cite{Goswami05}] of the canonical value were most
likely to involve contamination of a GMC having $\sim 3 \times
10^5$~\msun~of gas by the massive members of a cluster with $\sim
10^5$ stars.  Such a situtation is exemplified in our galaxy by NGC
3603, which contains multiple Wolf-Rayet stars within an HII region
\citep{Melena08}.

\section{Discussion \label{sec.discussion}}

\subsection{Delivery of \al26 into the molecular cloud \label{sec.efficiency}}

Our scenario requires a plausible mechanism for the efficient
introduction of the \al26 carrier in Wolf-Rayet winds into the
surrounding molecular cloud.  In the calculations above, we assumed a
100\% delivery efficiency but this will clearly not be the case.  In
general, mixing between the hot, tenuous gas from massive stars and
cooler, denser molecular gas is thought to be very inefficient
\citep{deAvillez02}.  SN-enriched gas from HII regions is thought to
find its way back into the interstellar medium (if at all) through a
circuitous route taking $\sim$100~Myr \citep{Tenorio00}.  Impact of SN
ejecta onto a protostellar cloud core may induce mixing via
Rayleigh-Taylor and Kelvin-Helmholtz instabilities
\citep{Foster98,Boss08} but the injection efficiency is too low to
explain the Solar System's canonical \al26/\a27
ratio\footnote{Adopting an \a27 mass fraction of $5.8 \times 10^{-5}$
  \citep{Lodders03}, the protosolar cloud core initially contained $3
  \times 10^{-9}$~\msun~of \al26.  The mass fraction of \al26 in the
  convective hydrogen shell of a 25~\msun~star at the end of its
  main-sequence life is $\sim 1 \times 10^{-6}$ \citep{Meyer05}.
  Therefore {\it at least} $3 \times 10^{-3}$~\msun~of SN ejecta must
  have been introduced (assuming no free decay).  Boss et al. (2008)
  report that only $5 \times 10^{-5}$~\msun~is injected in their
  model.}.  Gas in Wolf-Rayet winds will be less dense than SN ejecta
by several orders of magnitude and efficient mixing in the gas phase
is even less likely.  Instead, the high-velocity
(500-2000~km~sec$^{-1}$) winds will develop a reverse shock upon
encountering the much denser molecular cloud \citep{Weaver77}.

We propose that refractory dust grains are the principle carrier of
\al26 and \calcium41 in Wolf-Rayet winds and that these can
dynamically decouple from the shocked wind and imbed themselves into
the surrounding molecular cloud, analogous to the ``aerogel'' model
described by Ouellette et al. (2005).  Pre-solar grains of Al$_2$O$_3$
(corundum or other forms) are found in primitive meteorites but their
oxygen isotopes indicate a source in red giant or AGB stars, not in
very massive stars \citep{Hutcheon94,Ott07,Nittler08}.  Although many
of these grains have large $^{26}$Mg excesses produced by the decay of
of \al26, the abundance of these grains is insufficient to account for
the canonical \al26/\a27 ratio \citep{Hutcheon94}.  In fact, the \al26
in CAIs must have been processed through the gas phase and
subsequently recondensed into the refractory inclusions.  The
pre-solar grains are also much larger ($\sim$1~$\mu$m) than the
silicate grains predicted to form from winds and ejecta.  Their size
may be why they survived the incorporation process and the latter did
not.

Simple models of grain nucleation and growth predict oxide grain
growth to sizes of 0.01-0.1 $\mu$m in SN ejecta \citep{Nozawa07}.  SN
are predicted to be copious sources of dust, but observations have so
far produced evidence only for a few times $10^{-5}$~\msun~of dust in
individual events \citep{Meikle07}.  Dust production in WR stars is
poorly investigated, although such stars appear to be minor
contributors to the over-all interstellar dust budget
\citep{Tielens05}.  Copious amorphous carbon dust is observed around
carbon-rich Wolf-Rayet CO stars and is thought to be the result of
colliding stellar winds in binary systems \citep{Crowther03}.  WCO
stars do {\it not} produce \al26, but dust production by predecessor
LBV and WN phases predicted to contain \al26 in their winds has
recently been established \citep{Rajagopal07,Barniske08}.  The $\eta$
Carinae LBV star ejected $\sim$10~\msun of dust-rich materal during
its 1843 eruption \citep{Smith03}, including aluminum oxide
\citep{deKoter05}, although its \al26 content has yet to be definitely
established by $\gamma$-ray observations \citep{Knodlseder96}.

To reach the molecular cloud, grains must survive sputtering after
they pass the reverse shock and move at high-speed
($\sim10^3$~km~sec$^{-1}$) with respect to the shocked gas in the HII
region.  They also must not be completely decelerated within the
reverse shock zone.  These conditions place a lower limit on grain
size.  Grains too small are sputtered to destruction or eventually
vaporized in the hot, shocked gas \citep{Nozawa07}.  The density of
the wind 1~pc from a Wolf-Rayet star losing mass at a rate of
$10^{-5}$~\msun~yr$^{-1}$ is $\sim10^{-2}$~cm$^{-3}$ and the
stopping distance of grains even as small as 0.01~$\mu$m grains is
10~pc, comparable or larger to the size of HII regions.  The
deceleration across the scale of the shocked region ($\sim$1~pc)
\citep{Weaver77} will be low and the fraction of material sputtered
from the grains, which is related to the deceleration [Eqn. 1 in
Nozawa et al. (2007)] will be likewise small\footnote{In contrast, the
density of shocked SN ejecta is $\sim10^5$~cm$^{-3}$, the stopping
distance of a 0.1~$\mu$m grain is only $\sim$2~AU, whereas the ejecta
scale length can be as large as 1~pc.  Thus grains in SN ejecta are
more likely to be trapped in the ejecta and never introduced into
star-forming molecular gas.  This is another argument for WR winds as
the source of \al26 in the Solar System.}.

Grains that escape the WR wind will not penetrate far into a GMC.
Typical hydrogen number densities in clouds are $10^2-10^3$~cm$^{-3}$
and the stopping distance of 0.01-0.1~$\mu$m grains will be only of
order $\sim10^3$~AU.  Gas densities in portions of the molecular cloud
that are shocked and swept up by the expanding wind or SN ejecta will
be higher and the stopping distances proportionally shorter.  Thus,
only the surfaces of clouds will be initially contaminated by \al26.
Further transport of dust grains into cloud depends on their
kinematics, which are poorly understood.  The smallest scale on which
turbulence in clouds can affect the mass distribution and cause mixing
is the sonic transition where the turbulent velocity is equal to the
sound speed; this is roughly 1~pc in solar-metallicity clouds
\citep{Padoan95}.  Thus mixing might be inefficient at the surfaces of
clouds.  The degree to which this controls the incorporation of \al26
into new low-mass stars depends on the extent of large-scale mixing
and whether star formation is triggered or at least assisted, by the
interaction of winds or SN ejecta with cloud gas \citep{Zavagno07}.
In that case, star formation is spatially correlated with \al26
abundance.

Wolf-Rayet progenitors may themselves migrate into and contaminate
regions of a molecular cloud where low-mass star formation has yet to
occur.  10-30\% of O stars have large peculiar velocities (up to 200
km~sec$^{-1}$) relative to most early-type stars as a result either of
dissolution of binary systems by SN explosions, or encounters between
two binary systems \citep{Hoogerwerf01}.  One of these so-called
``runaway'' O stars moving at a typical speed of 30 km~sec$^{-1}$
would traverse a molecular cloud in $\sim$1~Myr.  Mass loss from this
star, if occuring, could contaminate a larger region of the molecular
cloud with SLRs.

\subsection{CAIs with low initial \al26/\a27 ratios \label{sec.funcais}}

Any scenario that explains the canonical Solar System \al26/\a27 ratio
must also accommodate the exceptions.  Several classes of CAIs show
either no excess of \mg26 produced by the decay of \al26 or have an
inferred \al26/\a27 ratio $\ll 1 \times 10^{-5}$, much lower than the
canonical value of (4.5-5)$\times10^{-5}$.  These include ({\it i})
igneous CAIs associated with chondrule-like materials (relict CAIs
inside chondrules and CAIs surrounded by chondrule-like,
ferromagnesian silicate rims) \citep{Krot05a,Krot05b}; ({\it ii}) some
igneous CAIs in metal-rich (CB and CH) carbonaceous chondrites
\citep{Gounelleetal07,Krot08b}; ({\it iii}) FUN (fractionation and
unidentified nuclear effects) CAIs \citep{Lee88}, ({\it iv}) isolated
platy hibonite crystals (PLACs) \citep{Ireland00}, ({\it v})
pyroxene-hibonite spherules \citep{Ireland91,Russell98}; ({\it vi})
some corundum-bearing CAIs \citep{Simon02}, and ({\it vii}) most of
the grossite- and hibonite-rich inclusions in CH chondrites
\citep{Kimura93,Weber95,Krot08b}.

({\it i} and {\it ii}): CAIs associated with chondrule materials and
some igneous CAIs in CB and CH carbonaceous chondrites (Krot et
al. 2001, 2005a,b, 2008b) are \o16-depleted to varying degrees
($\Delta^{17}$O ranges from -25\permil~to -5\permil) relative to
typical CAIs in primitive chondrites which uniformly have
$^{16}$O-rich compositions ($\Delta^{17}$O $\sim$ -25\permil)
\citep{Itoh04,Makide08}.  We infer that the \al26-poor and
\o16-depleted CAIs experienced late-stage melting and oxygen isotope
exchange, probably during chondrule formation, which could have reset
their \al26-\mg26 systematics.

({\it iii}-{\it vii}): The lower than the canonical \al26/\a27 ratio
in FUN CAIs, PLACs, pyroxene-hibonite spherules, some corundum-bearing
CAIs, and most of the grossite- and hibonite-rich inclusions in CH
chondrites can be explained by (a) their late formation, after decay
of \al26, (b) their early formation, prior to introduction of \al26,
or (c) the lack of the canonical budget of \al26 in their precursors.
Most of these CAIs have \o16-rich compositions
\citep{Goswami01,Simon02,Krot08b,Krot08c}, indistinguishable from
those of typical CAIs with the canonical \al26/\a27 ratio.  The rapid
evolution of the oxygen isotopic composition of the inner Solar System
\citep{Krot05c,Aleon07} and the short ($<10^5$~yr) duration of CAI
formation \citep{Thrane06,Jacobsen08} thus make (a) unlikely.  Similar
arguments can be used against (b).  Although an early formation is
possible if complete melting and exchange of oxygen isotopes occured,
this seems unlikely considering evidence for incomplete melting of
CAIs \citep{MacPherson05}.

We infer that the CAIs of categories {\it iii-vii} formed
contemporaneously with \al26-rich inclusions. The absence of canonical
\al26 in their precursors suggests either they formed prior to
homogenization of \al26 in the Solar System
\citep{Sahijpal98,Krot08b,Krot08c} or preferential loss of the
(uniformly distributed) \al26 carrier during thermal processing of the
CAI precursors \citep{Wood96}. Both explanations can be reconciled
with the presence of nucleosynthetic anomalies in some of these CAIs
\citep{Lee98} if the \al26 carrier contributed a distinct component to
the Solar System's stable isotope composition \citep{Lee98}. The second
explanation is more speculative, however. It hypothesises that (a) the
precursors of these CAIs was isotopically heterogeneous and, contrary
to typical refractory inclusions, escaped a cycle of complete
evaporation-condensation, (b) the carrier of \al26 was relatively
volatile, and (c) it was lost to varying degrees by sublimation of
these CAI precursors prior to their melting.  Although these
inclusions can be used as an evidence for heterogeneous distribution
of \al26 among CAI precursors, the scale of any heterogeneity was
probably limited because such inclusions are rare relative to
\al26-rich CAIs.

\subsection{Other SLRs \label{sec.others}}

{\bf \calcium41}: A further test of the wind model is whether it can
also reproduce the inferred initial \calcium41/\ca40 ratio of $1.5
\times 10^{-8}$ \citep{Sahijpal98}.  Published calculations of
\calcium41 yields in winds from massive stars are limited.
We considered a 60~\msun~progenitor for which \calcium41 and \al26
yields from the Wolf-Rayet winds were separately published
\citep{Arnould06,Limongi06}.  We accounted for the additional free
decay of \al26, which is ejected during the hydrogen-burning WN
Wolf-Rayet phase, while \calcium41 is produced in the later core
He-burning WCO phase $1.9 \times 10^5$~yr later.  The corrected ratio
of \calcium41 to \al26 in the ejecta is $\sim25$ times higher than in
the Solar System, but would be consistent if an additional time
$\Delta \approx 0.5$~Myr elapsed before CAI formation.  It is
interesting that this is approximately the same duration as the
Wolf-Rayet phase itself before the final SN Ib/c.

\begin{figure}
\epsscale{1}
\plotone{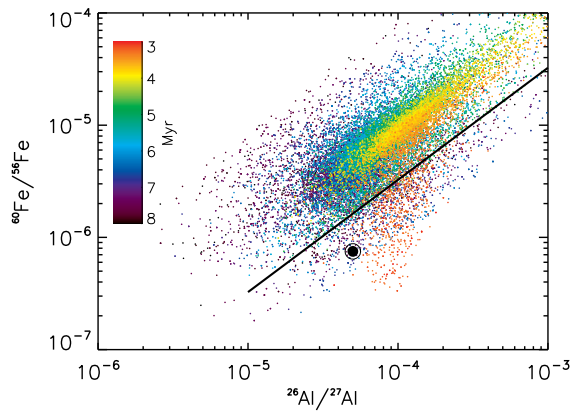}
\caption{Calculated abundances of \al26 and \fe60 relative to stable
comparison isotopes in star-forming regions contaminated by Wolf-Rayet
winds and SN ejecta (see text for details).  Each point represents a
Monte Carlo calculation of the composition of the gas in a well-mixed
molecular cloud 3-8 Myr after massive star fromation.  The large black
point is the inferred composition of the Solar System.  The line
represents the Galactic average abundance ratio from $\gamma$-ray
observations \citep{Wang07}.}
\label{fig.alfe}
\end{figure}

{\bf \fe60}: Wolf-Rayet winds contain negligible amounts of \fe60
\citep{Arnould06}.  Live \fe60 in the early Solar System could have
originated in SN from the same early generation of massive stars that
produced the \al26, or in an even earlier generation of stars
\citep{Gounelle08a}. We repeated the calculations described in \S
\ref{sec.cloud} but included SN contributions and calculated
\fe60/\f56 ratios in the same manner as \al26/\a27, using the yields
of Limongi \& Chieffi (2006) and the solar iron abundance of Lodders
(2003).  We assume $\Delta = 0.5$~Myr based on the \calcium41
abundance.  In Figure \ref{fig.alfe} we plot Monte Carlo realizations
for different epochs (3-8~Myr before the Sun) for the earlier
generation of massive stars.  If all SN ejecta is incorporated, the
\fe60/\f56 is over-predicted by an order of magnitude relative to
\al26/\a27.  A comparision with the ratio of the two SLRs (black line)
inferred from $\gamma$-ray measurements \citep{Wang07} suggests that
this discrepancy may be in part the result of an overprediction of
\fe60 yield - or underprediction of \al26 yield - by the
nucleosynthesis models (see \S \ref{sec.summary}).  There are two
other explanations suggested by Figure \ref{fig.alfe}: (a) the Sun
formed 3~Myr after the massive stars, when many massive stars were in
the WR phase but few SN had occurred (a scenario represented by the
red dots extending below the primary locus); or (b) the SN
contribution was attenuated by an effect such as described in \S
\ref{sec.efficiency}.

Explanation (a) is statistically unlikely if star formation is
uncorrelated and demands precise timing between the formation of
massive stars and the Sun ($\sim$3~Myr later), but could be demanded
in a scenario where the Sun's formation was triggered by a SN
\citep{Cameron77,Boss08}.  Our simulations indicate that the initial
solar \al26/\a27 and \fe60/\f56 ratios can be reproduced in this
manner only in star clusters with $N_* > 10^{5}$ whose most massive
members have $\sim100$ \msun.  As such large clusters are relatively
rare, this scenario is {\it a priori} less likely.  Explanation (b)
requires a reduction in the SN contribution to \fe60 (and \al26) by a
factor of 20.  This could be due to a combination of effects;
retention or fallback of the central region of the progenitor
\citep{Meyer05}, inefficient delivery of SN ejecta into the cloud (\S
\ref{sec.efficiency}) or the collapse of the protosolar cloud and a
decrease in its cross-section by the time the SN ejecta arrived.  An
alternative scenario (c) is that \fe60 in the Solar System is the
relict of an even earlier episodes of massive star-formation and
contamination \citep{Gounelle08a} of which the \calcium41 and most
\al26 has decayed.  \fe60 will decay to 5\% of its initial abundance
in 6.5~Myr, during which \al26 decays to 0.2\%, and \calcium41
essentially vanishes.  This last explanation is viable only if this
earliest generation of stars ceased to contribute SLRs to the
molecular cloud after $\sim$6~Myr.

{\bf \cl36}: We estimated the abundance of \cl36, which is also
ejected by WR stars during the WCO phase \citep{Arnould06}, and
calculated the \cl36.\chl35 ratio in the same manner as
\calcium41/\ca40.  We find that our model underpredicts the ratio by
at least three orders of magnitude.  It has already been recognized
that stellar nucleosynthetic models cannot account for this isotope,
especially if it was introduced at the epoch of CAI formation 1-2~Myr
before the host sodalite alteration phases were formed.  At the present
time, the only viable explanation appears to be a late episode of
irradiation by energetic particles \citep{Hsu06}

\section{Summary and Outlook \label{sec.outlook} \label{sec.summary}}

The canonical abundance of \al26 in the Solar System cannot be
explained in terms of a late injection of debris from a nearby SN into
the gaseous protoplanetary disk because ({\it i}) the dispersal of the
natal cluster and the finite time window for injection make it {\it a
  priori} an unlikely event ($<$2\%), ({\it ii}) \al26 was already
present in CAIs, which formed within $\sim10^5$~yr of the initial
collapse of the protosolar nebula and the formation of the
protoplanetary disk, and ({\it iii}) the oxygen isotope systematics of
primitive Solar System materials show no sign of a late introduction
of SN ejecta.  The apparently uniform distribution of \al26 in
meteorites and samples of the Earth, Moon, and Mars suggests
homogenization during the collapse of the protosolar cloud.

We showed that the canonical \al26/\a27 ratio can be explained if the
Solar System formed from a molecular cloud contaminated by Wolf-Rayet
winds from massive stars that formed 4-5~Myr earlier.  A SN
contribution is not required to explain the abundance of \al26,
although it is not necessarily excluded.  The {\it a priori}
probability that such a level of contamination occured depends on the
poorly-understood star formation histories in GMCs; we estimate that
it is $\sim$6\%.  However, our model predicts that the canonical value
is close to the {\it most likely non-zero value} in the Galactic
distribution.

The initial \calcium41/\ca40 ratio can also be explained by Wolf-Rayet
wind contamination if $\sim$0.5~Myr elapsed between its introduction
by winds and the formation of CAIs.  If this scenario is also to
explain primordial \fe60 in the Solar System, the cloud must have been
contaminated with SN ejecta as well.  If SN ejecta is included, our
model over-predicts the abundance of \fe60 by an order of magnitude.
This discrepancy could be rectified by some combination of the
following: (a) most \fe60 falls back onto SN remnants rather than be
ejected \citep{Meyer05,Takigawa08}; (b) most dust grains in SN ejecta
are retained and destroyed in the shocked ejecta and never enter the
molecular cloud; (c) the protosolar cloud was already collapsing and
presented a smaller cross-section when the SN ejecta arrived; and (d)
the \fe60 is a relict of an even earlier episode of massive star
formation and contamination for which all the other SLRs have decayed
away.  Absence of a significant excess of \mn53 seems to require (a),
but not to the exclusion of the other explanations.  Our model does
{\it not} explain the inferred abundance of \cl36, and another
mechanism such as irradiation much be invoked.

A key uncertainty in our model is the efficiency with which \al26 is
introduced into the host molecular cloud and the degree to which it
becomes uniformly mixed.  We propose that the carrier of \al26 was
dust grains and that these dynamically decoupled from the wind and
embedded themselves (intact) into the cloud, but this hypothesis needs
further investigation.  Furthermore, our model does not account for
the inhomogeneities in SN ejecta and WR winds that could produce
spatial variation in the contamination of a molecular cloud.  There
are also uncertainties in calculations of the evolution and
nucleosynthesis of massive stars that could quantitatively alter our
results.  Production of \al26 by neon burning during the Type Ib/c SN
that follows the Wolf-Rayet phase is sensitive to the progenitor mass
\cite{Higdon04} and for a 60 \msun~progenitor, could be as large as
the yield from the wind \cite{Woosley07}.  New models that include
stellar rotation predict higher yields of \al26, an earlier appearance
of \al26 in the WR wind (as early as 1~Myr), and a smaller minimum
initial mass for entry into the WR phase \citep{Palacios05}.  Larger
\al26 yields would relieve the requirement for high delivery
efficiency to the molecular cloud and may resolve the discrepancy
between the predicted amount of concomitant \fe60 from SN ejecta and
the inferred initial abundance of the radionuclide in the Solar
System.  Future tests of this model could compare predicted WR stellar
contamination with short-lived isotopes (e.g., \pd107) whose
abundances seem consistent with models of the interstellar medium
(G. Huss, pers. comm.), as well as the Solar System's oxygen isotopic
composition.

\acknowledgments

This material is based upon work supported by the National Aeronautics
and Space Administration through the NASA Astrobiology Institute under
Cooperative Agreement No. NNA04CC08A issued through the Office of
Space Science.  SR is a NASA Postdoctoral Program Fellow.  Some of
this work was performed while EG was a Visiting Scholar at the
University of California Berkeley.  We thank Gary Huss and Kazuhide
Nagashima for enlightening discussions and John Bally and a second,
anonymous reviewer for helpful comments and corrections.

\clearpage

\begin{deluxetable}{lrr|r}
\tablecaption{Statistics of \al26 abundance in different scenarios \label{tab.stats}}
\tablewidth{0pt}
\tablehead{
\colhead{Scenario} & \multicolumn{2}{c}{\%} & \multicolumn{1}{c}{\al26/\a27}\\
 & \colhead{$ > 0$} & \colhead{$ \geq 5 \times10^{-5}$} & \colhead{95\%}
}
\startdata
\multicolumn{4}{l}{Disk injection:}\\
T = 0.5~Myr & 16 & 1.2 &  $5 \times 10^{-6}$\\
T = 1~Myr & 20 & 1.9 & $1.3 \times 10^{-5}$\\
\tableline
\multicolumn{4}{l}{Disk injection with Bondi-Hoyle accretion}\\
T = 0.5~Myr & 16 & 1.2 &  $8 \times 10^{-6}$\\
T = 1~Myr & 21  & 1.9 &  $1.6 \times 10^{-5}$\\
\tableline
\multicolumn{4}{l}{Molecular gas contamination (T=0):}\\
2.7~Myr exp. SF & 16 & 6.2 &  $9 \times 10^{-5}$\\
1.7~Myr exp. SF & 8 & 4.3 &  $6 \times 10^{-5}$\\
3.7~Myr exp. SF & 21 & 6.5 &  $9 \times 10^{-5}$\\
\tableline
\enddata
\tablecomments{Only simulations which produced \al26-contaminated systems are reported.}
\end{deluxetable}

\clearpage

\end{document}